**RESEARCH ARTICLE**

# Unmasking the role of remote sensors in comfort, energy, and demand response

Ozan Baris Mulayim[1] , Edson Severnini[2] and Mario Bergés[1]

[1]College of Engineering, Carnegie Mellon University, Pittsburgh, PA, USA
[2]Heinz College, Carnegie Mellon University, Pittsburgh, PA, USA
**Corresponding author:** Ozan Baris Mulayim; Email: omulayim@andrew.cmu.edu

## Abstract

In single-zone multi-node systems (SZMNSs), temperature controls rely on a single probe near the thermostat, resulting in temperature discrepancies that cause thermal discomfort and energy waste. Augmenting smart thermostats (STs) with per-room sensors has gained acceptance by major ST manufacturers. This paper leverages additional sensory information to empirically characterize the services provided by buildings, including thermal comfort, energy efficiency, and demand response (DR). Utilizing room-level time-series data from 1000 houses, metadata from 110,000 houses across the United States, and data from two real-world testbeds, we examine the limitations of SZMNSs and explore the potential of remote sensors. We discover that comfortable DR durations (CDRDs) for rooms are typically 70% longer or 40% shorter than for the room with the thermostat. When averaging, rooms at the control temperature's bounds are typically deviated around $-3$ °F to 2.5 °F from the average. Moreover, in 95% of houses, we identified rooms experiencing notably higher solar gains compared to the rest of the rooms, while 85% and 70% of houses demonstrated lower heat input and poor insulation, respectively. Lastly, it became evident that the consumption of cooling energy escalates with the increase in the number of sensors, whereas heating usage experiences fluctuations ranging from $-19\%$ to $+25\%$. This study serves as a benchmark for assessing the thermal comfort and DR services in the existing housing stock, while also highlighting the energy efficiency impacts of sensing technologies. Our approach sets the stage for more granular, precise control strategies of SZMNSs.

**Impact Statement**

This study challenges assumptions made about single-zone multi-node systems (SZMNSs), such as the belief that there are no significant temperature discrepancies within them and that thermostats are located in places that would represent the overall house dynamics. Through empirical evidence, we (a) demonstrate the severity of temperature variations in SZMNSs, (b) evaluate the effectiveness of averaging strategies using sensor networks, (c) diagnose the root causes of such variations by using gray box modeling approaches, and (d) analyze the energy implications of having additional sensors. Our analysis regarding the prevalence of additional sensors supports the potential of using additional sensory information to evaluate the services provided by buildings such as thermal comfort, energy efficiency, and demand response.

## 1. Introduction

Commercial Heating, Ventilation, and Air Conditioning (HVAC) systems enable zonal controls, facilitating temperature optimization in each respective zone using a variety of control solutions. However, in residential settings, the heating and cooling mechanisms are typically dictated by a single temperature probe that oversees the entire building's thermal regulation due to the cost of multi-zone systems (Jornet-Monteverde and Galiana-Merino, 2020). This invariably culminates in variances in room temperatures, adversely impacting occupant thermal comfort (Lin et al., 2002; Seri et al., 2021) and the quality of other services provided by these buildings such as demand response (DR). This issue is particularly accentuated in multi-floor dwellings, a prevalent residential layout in the United States.

Although few studies have empirically identified temperature variations (Lin et al., 2002; Shen et al., 2023), a comprehensive analysis, especially one involving a larger sample size, has yet to be undertaken. Notwithstanding, despite not having sufficient evidence of how severe the deviations are, research has proposed numerous solutions, such as automatic vent registers (Redfern et al., 2007; Sookoor and Whitehouse, 2013; Jornet-Monteverde and Galiana-Merino, 2020; Shen et al., 2023), remotely controlled space heaters (Klingensmith et al., 2014), and averaging strategies (Lin et al., 2002). Companies like Keen Home (2023) and Alea Labs (2023) have produced smart vents, but adoption is limited due to the need for complex installation processes and waste production (Forest and Shabani, 2017). On the other hand, averaging strategies offered by major smart thermostat (ST) companies, such as *ecobee* (2024a) and Nest (2023), have a higher likelihood of adoption. This is predominantly because STs already have a substantial market presence in the United States, with an adoption rate of 11% (King, 2018). However, the effectiveness of these averaging methodologies has not been thoroughly investigated. Additionally, leveraging the data from STs, especially with the advent of multi-sensor setups, offers an opportunity to shed light on the root causes of such discrepancies.

In this study, we analyze multiroom sensing data derived from a large sample of STs in the United States and two single-family residential test beds. The focus of these inquiries is to **evaluate the extent** of temperature variations, **gauge the efficacy** of commercially available thermostats, **diagnose the limitations** contributing to these deviations, and **assess the energy implications** of remote sensors. Specifically, the main contributions of this paper are (1) analyzing the prevalence of remote sensors and DR participation among *ecobee* users across a dataset of around 110,000 houses, (2) assessing temperature deviations during operational conditions and DR events, (3) examining the benefits and limitations of averaging methodologies through an experimental study and a large dataset encompassing approximately 900 houses, (4) identifying the underlying causes of thermal variations by performing a thermal characterization of up to 100 houses, equipped with five additional sensors, and (5) conducting statistical analysis on yearlong data from 1000 houses to identify the impact of having additional sensors on energy consumption. Our findings provide a valuable benchmark in evaluating the quality of services such as thermal comfort and DR provided by the existing housing stock. They also shed light on the need for room-level retrofitting, and energy consumption impacts of remote sensors, thereby promoting enhanced energy efficiency. Ultimately, our empirical analysis lays the foundation for more granular, precise control strategies utilizing sensor networks for residential HVAC systems.

This paper builds upon our previous research (Mulayim and Bergés, 2023), which focused on analyzing thermal behavior through the use of remote sensors. In the current study, we have expanded our analysis to include metadata from approximately 110,000 households equipped with *ecobee* systems. This extension allows us to assess the penetration rate of remote sensors, detailed in Section 4. Furthermore, we explore the participation rates in DR programs through *ecobee* and focus on households' preferences for setpoint changes during DR events, as well as the availability of remote sensors within these participating households in Subsection 5.2.1. Finally, we elucidate the impact of remote sensors on energy consumption across 1000 households through an explanatory data analysis and a comprehensive statistical analysis presented in Section 8. This expansion of our research scope not only offers a deeper insight into the role of remote sensors in managing energy consumption but also highlights their widespread availability, suggesting their potential inclusion in future studies aimed at evaluating services provided by buildings.

The remainder of the paper is structured as follows: Section 2 reviews related literature, followed by an explanation of our methodologies and datasets in Section 3. We then analyze the prevalence of remote sensors in Section 4, identify temperature variations in Section 5, evaluate averaging techniques in Section 6, diagnose potential factors leading to these variations (Section 7), and assess the energy implications of remote sensors in Section 8. The paper concludes with a discussion in Section 9.

## 2. Literature review

We divide the literature review into two primary focus areas. First, we consider studies that delineate the challenges of SZMNSs and/or propose enhancements. Our examination of the initial set of literature reveals a lack of studies that utilize a large sample size to empirically analyze the intensity of these variations and identify their root causes. That is why we conducted a second literature search where we investigated the use of *ecobee* datasets to study the thermal attributes of buildings. We believe that the methods used in this area of research could be deployed in a room-level granularity to identify the reasons for deviations considering existing housing stock. Though there is an intersection of these two literatures, to the best of our knowledge there are no studies identifying room-level thermodynamic models of SZMNSs.

### 2.1. Single-zone multi-node systems

Studies conducted on typical Canadian homes revealed that temperatures fell within a comfortable range (±1 degree Celsius from thermostat reading) about 40% of the time, indicating a substantial proportion of time spent in uncomfortable thermal conditions (Shen et al., 2023). However, this may be underestimating the problem given that it disregards the possible deviations that might occur from the setpoint by the thermostat reading itself.

Some studies (Redfern et al., 2007; Sookoor and Whitehouse, 2013; Jornet-Monteverde and Galiana-Merino, 2020; Shen et al., 2023) and commercial ventures (Alea Labs, 2023; Keen Home, 2023) have proposed multi-zone control enabled by automatic vent registers as a solution. However, these techniques may present substantial risks, such as potential damage to ducts and motors due to overpressure accumulation (Sookoor and Whitehouse, 2013; Jornet-Monteverde and Galiana-Merino, 2020). Additionally, closing registers has been shown to cause more leaks in the ducts, which reduces the efficiency of the system (Walker, 2003). Moreover, these techniques do not scale efficiently to multistory buildings (Klingensmith et al., 2014).

Further investigations have assessed the efficacy of supplemental equipment, like register booster fans and space heaters, in tandem with multi-nodal sensing controllers (Klingensmith et al., 2014). Register booster fans were found to be ineffective in rooms with inadequate insulation or a significant distance from the furnace. In contrast, although space heaters were proven to be useful, they pose challenges in terms of installation costs and aesthetic integration.

The potential utility of sensor networks for enhanced control of SZMNSs was initially explored by Lin et al. (2002), who conducted a simulation analysis of various strategies based on either averaging room temperatures or prioritizing the worst-performing room. Additionally, several studies have revealed a prevalent trend among occupants of multistory homes to condition their entire dwelling to maximize comfort in one or two rooms, leading to substantial energy waste (Redfern et al., 2007; Klingensmith et al., 2014). This demonstrates that a more homogeneous temperature distribution among the house would also result in improved energy savings, as shown in Lin et al. (2002).

### 2.2. Building thermal characterization

Numerous studies have worked on *ecobee*'s Donate Your Data (DYD) (*ecobee*, 2023b) datasets for building thermal parameter estimation. In John et al. (2018), an *ecobee* dataset, comprising 10,000 houses with STs, was utilized to predict thermal time constants (TTCs). Their findings revealed notable disparities in TTCs between the summer and winter months. Thermal building parameters, as well as TTCs, were identified by utilizing winter months' data from 4000 houses in Ontario and New York by Baasch et al. (2019). The work presented in Huchuk et al. (2022) utilized data from 1000 houses within the *ecobee* dataset to train various data-driven models, including a gray box model featuring indoor temperature, outdoor temperature, HVAC run time, and solar irradiance. Nevertheless, no analysis relating to the thermal parameters of the buildings was conducted. Among the studies reviewed, Wang et al. (2021) bear the closest resemblance to our research in terms of TTC identification. They conducted a dual parameter identification analysis during heating and cooling seasons to infer TTCs and equivalent temperatures due to solar and internal heat gains. As the primary aim of this study was to create a simulation environment for control algorithms, the detailed calculations were not further elaborated upon.

Given the shortcomings of current methodologies for improving comfort in SZMNSs, there is an imperative need for the implementation of more straightforward, scalable, and safe strategies to enhance comfort in SZMNSS. Preliminary successes of averaging methodologies have been documented in a limited number of simulation studies (Lin et al., 2002), and these techniques have been adopted by ST enterprises. Nonetheless, several important knowledge gaps remain: the severity of the temperature variations, the efficiency of these averaging strategies, and the underlying reasons for the limitations that cannot be addressed merely by averaging.

## 3. Methodology

In this section, we first delve into the metrics adopted for quantifying the successful operation of buildings, providing a comprehensive understanding of the measures we utilize. Following this, we outline the parameter identification methodology applied to extract the thermal parameters of rooms. This procedure is essential for diagnosing limitations inherent in the existing housing stock. Lastly, we present an overview of the datasets utilized throughout the paper, illustrating the breadth and depth of empirical evidence underpinning our study.

### 3.1. Comfortable operation metric

Various metrics have been deployed to evaluate the comfortable operation of buildings. Percentage indices have been used to evaluate the frequency of temperature deviations outside the comfort range (Seri et al., 2021; Shen et al., 2023), while risk indices measure human thermal comfort perception (Carlucci and Pagliano, 2012). Cumulative indices are the only kind that considers both frequency and the magnitude of the deviation together. However, most of them require complex parameters and detailed knowledge about occupant characteristics. For instance, $Exceedance_M$ is an asymmetric discomfort index that sums over occupied hours of overheating (Borgeson and Brager, 2011), but its use often demands occupancy data, which many residences lack. Degree-hours criterion defined by ISO 7730 (International Organization for Standardization, 2005) is easy to apply and does not require extensive knowledge like others, but it is a cumulative value making it harder to compare between buildings. Unmet load hours, though widely recognized, is sensitive to sample size and indiscriminately penalizes both minor and significant setpoint deviations.

The closest metric to our work was introduced in a recent study by Shen et al. (2023). Their method computes the relative frequency (RF) for each deviation interval (e.g., 2 to 3 °F) by dividing the number of times the room was in that interval by the length of the whole dataset. Although this is a useful method that demonstrates how often a value was in a certain deviation interval from the thermostat reading, it fails to capture the general behavior of the room in two ways: (1) it does not consider discomfort as a deviation from user-requested setpoint; (2) it can only focus on a certain interval and thus does not produce generalizable results. These shortcomings suggest the need for a metric that can be used on a room-level resolution and consider both the magnitude and the frequency of the deviation from the setpoint, while still being limited to a certain scale (i.e., 0–1). In response to this methodological gap, we will be using a new metric: the Comfortable Operation Index (COI). This index, as shown below, measures how successful the HVAC is at conditioning each room, considering a specific temperature setpoint and deadband.

$$\text{COI} = \frac{\sum_{x=-C}^{C} R(x)}{\sum_{x=-M}^{M} R(x)} \tag{3.1}$$

$R(x)$ is the RF of the given temperature deviation interval $x$ from the temperature setpoint. The set $\{-C, \ldots, C\}$ represents the range of temperature deviations defined as within the *comfort zone*. The set $\{-M, \ldots, M\}$ is the temperature deviation limits of the analysis range (decided by the analyst). Using the maximum absolute deviation in the dataset as $M$ makes the denominator 1, converting the function to approximate a probability density function, as applied in this study. However, analysts might prefer limiting the range to different percentiles. For results consistent with ours, $M$ should equate the dataset's maximum deviation (denominator = 1) and $C$ should be set at 2 °F. Utilizing alternative $M$ and $C$ values confines comparisons to specific rooms or houses within a given study. In essence, COI represents the portion of the RF curve's area within the comfort range, compared to the total area for the analysis range (See Figure 3).

The COI has an inherent limitation: it requires a singular setpoint and a deadband, making it less applicable to HVAC systems operating with dual setpoints for heating and cooling. While the COI can adapt by taking the average of both setpoints in "auto" mode, this approach becomes misleading if users exclusively employ either heating or cooling modes. To address this, we introduce the Comfortable Cooling Index (CCI) for analyzing cooling operations. Unlike COI, CCI defines the *comfort zone* as $\{-M, \ldots, 0\}$ and calculates deviation relative to the cooling setpoint.

### 3.2. Parameter identification

#### 3.2.1. Free floating periods

Free floating periods (FFPs) have been successfully used to estimate gray box modeling parameters from ST data (John et al., 2018; Baasch et al., 2019; Wang et al., 2021). They can be described as periods where no additional heating or cooling occurs. Constraints defined for extracting FFPs are different for heating and cooling seasons. For the heating season, FFPs were constrained to occur between 10 p.m. and 7 a.m., given that the internal and solar gain during this period is negligible (Wang et al., 2021). Additionally, there needed to be a temperature difference of 2 °F for the sensor of focus, the FFP had to last at least an hour, and the outdoor temperature had to be lower than the indoor temperature. For the cooling season, in contrast, solar gain is not negligible. Hence, we considered the period where solar gain was high to infer the largest solar gain, akin to the approach taken by Wang et al. (2021). In this case, the FFP was expected to happen between 10 a.m. and 5 p.m. Similarly, there had to be a temperature difference of 2 °F for the sensor of focus, the FFP needed to last at least an hour, and the outdoor temperature had to be higher than the indoor temperature during free floating. Following Baasch et al. (2019), Wang et al. (2021), we are going to use a thermal Resistance-Capacitance ($RC$) modeling approach for each sensor using FFPs. The thermal $RC$ model for predicting indoor temperature dynamics is based on several simplifying assumptions for tractability and ease of analysis. It treats the space as a lumped parameter system with uniform temperature, employs first-order dynamics to describe thermal behavior, and assumes constant thermal resistance and capacitance values, implying linear heat transfer processes and neglecting material property variations with temperature. The model also assumes steady external conditions, particularly a constant mean outdoor temperature, overlooking potential fluctuations. These assumptions enable the model to offer a simplified yet effective framework for understanding and managing the thermal characteristics of indoor environments. The structure of the model can be found below.

$$\theta(t) = \theta(0) \cdot e^{-\frac{t}{RC}} \tag{3.2}$$

$$\theta(t) = T_i(t) - T_o - RQ \tag{3.3}$$

where $T_i(t)$ is the indoor air temperature of the sensor of focus at time $t$, $T_o$ is the mean outdoor temperature, $RQ$ is the thermal resistance of the room times heat input (in this case, solar and internal gain), $RC$ is the thermal resistance times capacitance (also referred to as TTC), and $\theta(t)$ (shown in Equation (3.3)) represents the temperature difference between the indoor air temperature at the sensor location and the outdoor environment. Equation (3.2) describes the temporal evolution of the temperature difference $\theta(t)$, considering the initial temperature difference $\theta(0)$ and the time constant $RC$.

For the analysis of the heating season, we made the assumption of no internal or solar heat gains during the FFP, leading to the exclusion of the $RQ$ term and inference of only the $RC$ value. However, in the cooling season analysis, both the $RQ$ and $RC$ terms were inferred. The `scipy.optimize` library and its `curve_fit` function were employed for parameter identification. After identification, outliers were eliminated by using two standard deviation ranges from mean and error filtering of Root Mean Square Error (RMSE) values that are larger than 1 °F.

### 3.2.2. Balance point

The balance point method (introduced by Baasch et al., 2019) focuses on the relation between the energy usage of a building and outdoor temperature. Certain constraints are applied to extract suitable periods of data for the deployment of this method. First, a selection criterion is applied to isolate the nights when the heating system was operational and no cooling events occurred. The focus on nighttime data serves to minimize the influence of additional heat gains, such as solar or internal gains. The specified night period extends from 10 PM to 7 AM. As some residences employ a two-stage heating system, the run times of both the first and second stages are combined into a single metric to maintain consistency across different households. Few households in the sample have second-stage heating, which further justifies this unification.

Subsequently, the refined dataset is subjected to the balance point method. The structure of the model is as follows (details of the model can be found in Baasch et al., 2019):

$$F_{h,d} = \frac{1}{RK}\left(\overline{T}_i - \overline{T}_o\right) \tag{3.4}$$

where $F_{h,d}$ is the heating duty cycle of the night, $\overline{T}_i$ is the average indoor air temperature of the sensor of focus at night, $\overline{T}_o$ is the average outdoor temperature at night, and $RK$ is the thermal resistance of the room times heating power. $RK$ itself is not easy to interpret, but it can be used to make comparisons between rooms to decide if certain rooms are achieving less heat input from the HVAC system. The `scipy.stats` library and its `linregress` function were employed for parameter identification. After identification, a two-step filtering process was undertaken. Firstly, R-values below 0.7 were excluded, and subsequently, values deviating beyond two standard deviations from the mean were removed.

### 3.3. Dataset

#### 3.3.1. Test bed 1

The first test bed of this study (from here on named as *Test bed 1*) was a two-story, six-room house in Pittsburgh, PA, outfitted with a conventional mercury thermometer and a central HVAC system. During the data collection period, both heating and cooling were active, reflecting seasonal transitions. The thermostat and Room 1 are located on the first floor, while the other rooms are on the second floor. Temperature probes, consistent with ASHRAE guidelines, were placed 130–160 cm above the floor. Although data were gathered over 17 days in operational conditions, only 200 hours were usable for analysis due to sensor disconnections.

#### 3.3.2. Test bed 2

In this test bed, an *ecobee* thermostat was installed in a four-story, ten-room house located in Pittsburgh, PA. Data were accumulated over a 31-day period. The preprocessing included two steps: (1) considering the instances when the thermostat setpoint was 78 °F and (2) extracting a period of time where the average outdoor temperature was almost equal for both averaging and thermostat-based control (i.e., 72.5 °F). This preprocessing led to a remaining operational period of 264 hours. It can be seen that data collection did not take place during a significantly hot period with a low setpoint; therefore, temperature variations might be more pronounced in the hotter summer months.

### 3.3.3. ecobee *DYD dataset*

This study utilizes the *ecobee* DYD dataset, which comprises data from 1000 houses across the United States for the year 2017. Comprehensive details of this dataset are elaborated in Luo and Hong (2022). The analysis undertaken is divided into heating and cooling seasons due to two primary reasons: first, *ecobee* maintains separate cooling and heating setpoints rather than a single one; second, RC values were observed to vary significantly between winter and summer months (John et al., 2018). We also used different heating and cooling season definitions for thermal comfort and parameter identification analysis.

For thermal comfort analysis, this study focuses on households possessing at least one additional sensor beyond the thermostat reading, yielding a total of 887 and 845 households included in the cooling and heating season analyses, respectively. The counts vary over the year as some households acquire additional sensors subsequently. The heating and cooling seasons for the thermal comfort analysis are defined separately for each household by identifying the longest period during which **only** heating or cooling was active.

In parameter identification analysis, we used houses with five additional sensors that resulted in up to 100 houses. Earlier parameter identification studies have predominantly focused on either the winter or summer months. However, our preliminary investigation suggests a marked improvement in the identification percentage of parameters when the entire heating or cooling season is taken into account. For instance, we define a heating season for each house as the period that commences with the first activation of the heating system during the nine-month window (i.e., excluding summer months) and concludes with the final deactivation of the heating system within the same period. The same methodology was conducted for cooling season. Our study revealed a substantial increase in the number of heating and cooling days compared to the conventional 90-day periods.

### 3.3.4. ecobee *metadata*

The metadata for *ecobee* consists of information from 112,983 houses that participate in the DYD program. This information is available publicly in Luo and Hong (2022). It includes details about the characteristics of the houses, such as their floor area, city and state of location, number of occupants, number of floors, cooling and heating equipment stages, and HVAC equipment type. Most of the attributes are entered manually by the users, which results in multiple instances with unreasonable values such as 0 square feet, 0 number of occupants, and so on. Therefore, preprocessing is done separately for each attribute to ensure valuable information is not lost.

## 4. Penetration rate of remote sensors

To better understand the potential applicability of the methods described in this paper, we must first examine how easily we can access homes that have remote sensors. Our analysis involves two steps. First, we use the *ecobee metadata* to determine the distribution of the number of remote sensors. Second, we examine the correlation between the number of remote sensors and various factors such as location, number of floors, number of occupants, and floor area.

Our analysis begins by examining the number of remote sensors available in homes registered to the DYD program as of 2022. We found that 75,648 out of a total of 112,983 homes, which represents 67%, have at least one additional sensor (as shown in Figure 1). We believe this high number is due to the newer *ecobee* models that come equipped with an additional remote sensor. Considering that certain *ecobee* thermostats (premium) come with an additional sensor, the only way to have two or more sensors is by ordering a pack of remote sensors. Therefore, we can conclude that 35,647 users, which represents 32%, ordered additional sensors. Since sensors are sold in packs of two, this may help explain the slightly higher prevalence of odd-numbered values in Figure 1. Overall, this analysis shows us that additional sensors are becoming more and more prevalent. Thus, we see that there is potential to utilize additional sensing points in 67% of the houses with STs.

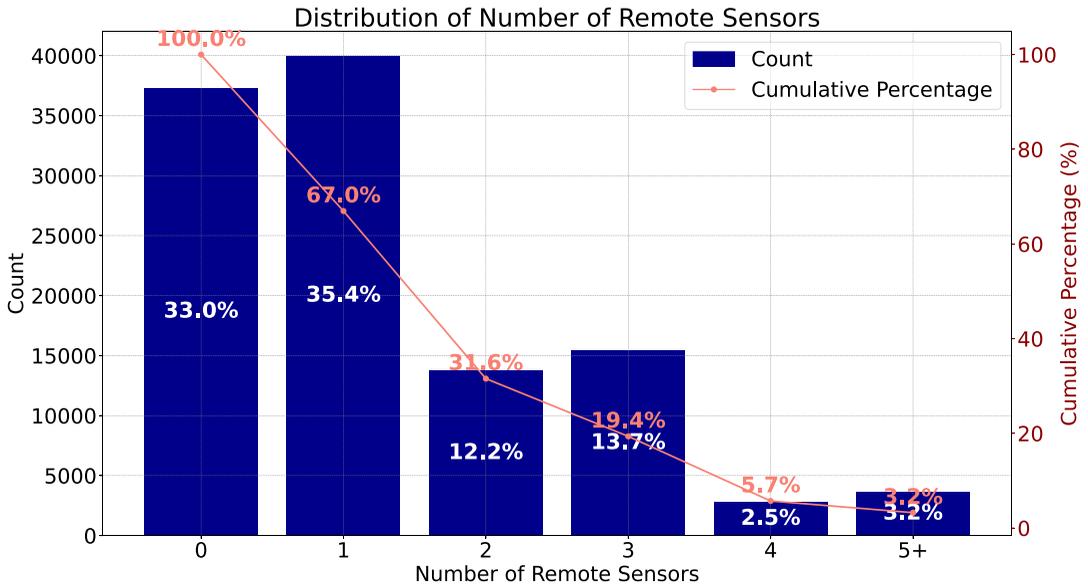

**Figure 1.** *Distribution of remote sensors across ecobee metadata. The vertical bars demonstrate the count of houses with a specific number of sensors, while the red line indicates the cumulative percentage, reflecting the proportion of the population with that many sensors or more.*

Through further analysis in Subsection 5.2.1, we demonstrate that among the 15,404 households participating in the eco + program, 11,253 (73%) are equipped with at least one sensor, and 5505 (36%) have two or more sensors. This finding suggests that for 73% of the households engaged in DR programs, there exists the potential to enhance thermal comfort assessment through the integration of additional sensor data.

In Figure 2, we explore the underlying motivations for households purchasing an increased number of sensors, considering factors such as the number of floors, occupants, and floor area. The top plot illustrates the distribution of homes equipped with various quantities of remote sensors, with each home's data distinguished by specific attributes. It is evident that the frequency of sensor deployment varies across attributes, a discrepancy attributed to the application of distinct filters during data processing to exclude entries with unrealistic values. Specifically, entries with zero occupants were omitted, and those with floor areas under 100 ft$^2$ were also discarded.

Our initial hypothesis posited a direct correlation between the proliferation of sensors and an increase in the number of occupants, floors, and overall floor area. A closer examination of the lower plots reveals that any decrease in the attribute proportion against an uptick in sensor numbers suggests a proportional relationship. Particularly, households with four or more occupants show a pronounced preference for having four or five or more sensors. Conversely, the increment in sensor possession among homes with three occupants and three or fewer sensors is minimal.

Analyzing the data based on the number of floors presents a clearer trend: as the number of sensors escalates, so does the proportion of homes with additional floors. It was noted that single-floor homes are least likely to have just two sensors, typically opting for more than five. Two-story homes constitute the largest share across all sensor categories, predominantly outfitted with two to four sensors. Residences exceeding three floors are more inclined to install a higher count of sensors, reflecting a positive correlation between floor count and sensor quantity.

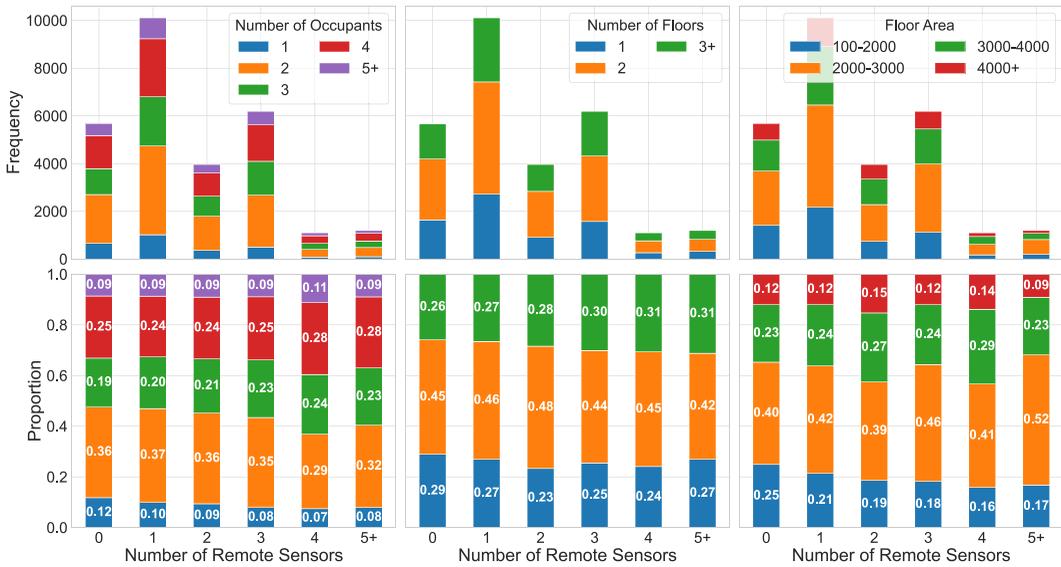

***Figure 2.*** *Distribution of remote sensors according to household characteristics: (a) number of occupants, (b) number of floors, and (c) total floor area. The top plots show the number of households in the dataset, while the bottom plots show their proportion within the chosen household characteristic.*

The analysis of floor area introduces a nuanced perspective: homes under 2000 square feet progressively demonstrate a lesser likelihood of adopting a greater number of sensors. This trend aligns with expectations, as smaller dwellings tend to exhibit less temperature variance. However, the data for larger homes do not consistently support a clear trend, displaying fluctuations in sensor numbers across different size categories. This variability hinders our ability to definitively conclude the relationship between larger homes (over 2000 square feet) and their propensity for sensor installation. Overall, a more systematic analysis (e.g., statistical regression) may be warranted to better understand these correlations.

## 5. Characterizing temperature variations

In this section, we assess the extent of temperature deviations during regular operation and DR events. Our analysis provides an understanding of discomfort in relation to setpoints, a factor previously overlooked in research. Furthermore, we explore the often disregarded thermal comfort implications of DR events (Li et al., 2021).

### 5.1. Variations with an existing thermostat

Figure 3 uses data from *Test bed 1* and the COI (explained in Section 3.1). On the vertical axis, RF values are plotted against each temperature deviation value, denoted as $x$, which ranges from 0—indicating that a specific deviation $x$ is never encountered—to 1, signifying that only the particular deviation $x$ is experienced throughout the entire duration of data collection. One should note that optimal operation of the thermostat would result in the peak location of the thermostat being located at 0 °F deviation. However, Figure 3 shows that the thermostat probe's measurement is persistently divergent by 3 °F, yielding a sustained fluctuation. Low COI values (shown in the legend) underscore the criticality of the situation for each individual room. We observe that the thermostat is not located in the most comfortable room, and the severity of discomfort increases as we move to rooms on the upper floors. While the deviation of the thermostat could be attributed to faulty installation of the mercury thermometer, the main reasons behind deviations in other rooms cannot be explained easily and need in-depth examination. For comparison, it is reasonable to assume that in a single-zone, single-node system, the thermostat line would reflect the entirety of the system given that control in *Test bed 1* is solely dictated by the thermostat temperature.

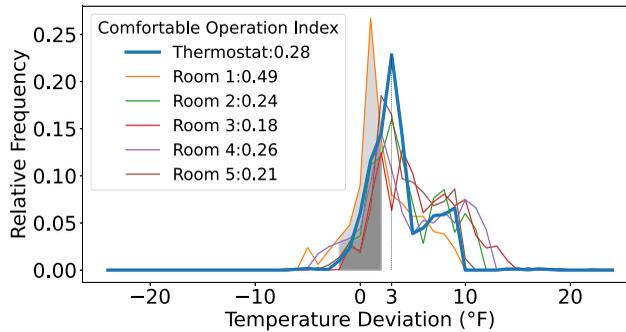

**Figure 3.** *Relative frequencies of temperature deviations from the setpoint in various rooms, represented by different colored lines. The gray area highlights the comfort zone, within which temperature fluctuations are considered comfortable for occupants. The COI values, detailed in the legend for each respective room, quantify the proportion of time temperatures were maintained within the comfort zone.*

The outcomes of this preliminary investigation not only emphasize the significance of assessing variations from the setpoint rather than the thermostat reading, but they also illustrate the gravity of scenarios wherein rooms frequently deviate from the comfort range.

## 5.2. DR effects

### 5.2.1. DR participation rate
In this subsection, we begin by analyzing the participation rate in DR programs, utilizing metadata from *ecobee*. We then examine how our testbed responded to a DR event. Finally, we evaluate the *ecobee* DYD dataset to confirm the presence of behavioral discrepancies among rooms during DR events, across larger sample sizes.

The analysis of *ecobee* metadata unveils the DR participation rate among 112,983 households, offering insights into homeowners' decisions regarding DR engagement and the prevalence of DR participants.

Central to this investigation is the eco + slider, a feature on *ecobee* thermostats that enables users to enroll in DR programs directly from their thermostat interface, provided such programs are available in their region. The level of the eco + slider represents the temperature precool/preheat and setback setpoint changes (in °F) that a user permits. Further details can be found in *ecobee* (2024b). Initial data analysis showed that 15,379 homes, or 13.6% of all households, have joined the eco + program.

Figure 4 illustrates the distribution of eco + slider level settings among users. Nearly half (44.9%) of the participants allow a precool/preheat adjustment of 4 °F and a setback temperature change of 2 °F in their temperature setpoint. Furthermore, about 80% of the users are comfortable with setpoint changes of 2 °F or more. This indicates a significant willingness among *ecobee* users to adapt their temperature settings, highlighting their flexibility and commitment to contributing to decarbonization efforts through partici- pation in DR programs.

According to the data, out of 15,404 houses that participate in eco+, 11,253 of them (73%) have at least one sensor, and 5505 houses (36%) have two or more sensors. This indicates that for 73% of the houses that participate in DR, there is potential for incorporating additional sensory information while assessing thermal comfort.

### 5.2.2. DR effects on Test bed 2
On June 19, 2023, we replicated a DR event at 12:00 pm on our entirely unoccupied *Test bed 2*, during which the average outdoor temperature was 85 °F. Figure 5a shows the comfortable DR duration (CDRD)

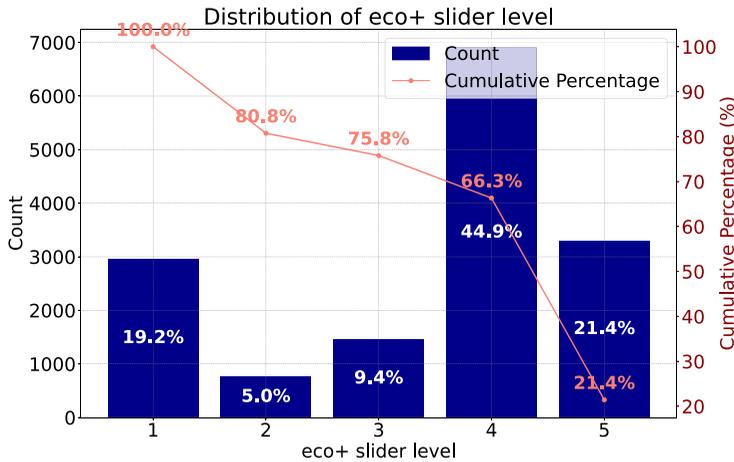

**Figure 4.** *The distribution of eco + slider levels among households participating in DR events through eco+. The bar chart indicates the number of households at each eco + slider level, while the red line traces the cumulative percentage, reflecting the proportion of households that have set their eco + slider to that level or higher.*

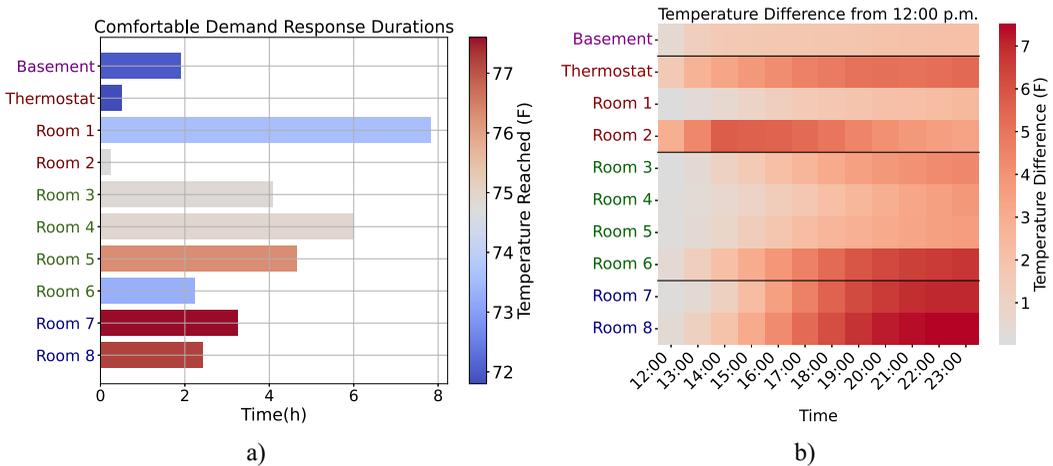

**Figure 5.** *A visual analysis of temperature shifts during a DR event, with the rooms distinctly color-coded and ordered on the y-axis to reflect their floor-level groupings. In Panel (a), a bar chart shows the elapsed time for temperatures to increase by 2 °F from noon across various rooms, while Panel (b) presents a heatmap depicting the temperature differential over time. Rooms situated on the same floor share color coding, and the sequence on the y-axis is arranged such that rooms on upper floors are positioned lower.*

(defined as the duration of time it takes for a room's temperature to rise by 2 °F in the absence of cooling) for each room colored by the final temperature reached, providing a detailed view of how a DR event can differently affect comfort levels in various rooms. For example, Room 1 could maintain a comfortable temperature throughout an 8-hour DR event, whereas Room 2 could only do so for around 15 minutes. Moreover, Figure 5b illustrates the temperature increase over a 12-hour span, beginning at 12:00 p.m. Rooms are categorized and color-coded based on their location within the structure, arranged in ascending order. The distinct time intervals seen in each room's temperature rise highlight the unique thermal behavior of individual spaces. For instance, Rooms 6, 7, and 8 reach their peak temperatures close to midnight, which may be a consequence of their specific orientation. Conversely, Room 2 experiences its peak temperature relatively early, likely attributable to its extensive window area. Room 1, on the other hand, displays a very small increase in temperature over a long period, which can be explained by its superior insulation.

**Table 1.** *Summary statistics of the CDRDs and temperature deviations of rooms during DR events*

| Parameter | Mean | Median | Std. dev. | Min | Max |
|---|---|---|---|---|---|
| **Durations (mins)** | | | | | |
| Fast-reacting rooms | 28 | 25 | 18 | 5 | 95 |
| Slow-reacting rooms | 80 | 69 | 35 | 40 | 230 |
| Thermostat | 47 | 40 | 31 | 5 | 150 |
| Comfort gap | 52 | 46 | 31 | 0 | 205 |
| **Deviations (°F)** | | | | | |
| Least-varying rooms | 1.92 | 2.15 | 1.27 | −3.00 | 4.05 |
| Most-varying rooms | 4.29 | 4.00 | 1.71 | −1.00 | 8.57 |
| Thermostat | 2.90 | 2.85 | 1.89 | −2.29 | 8.57 |
| Comfort gap | 2.37 | 2.00 | 1.49 | 0.00 | 6.14 |

### 5.2.3. DR effects on ecobee DYD dataset

In order to validate that these discrepancies are not isolated to a single household but persist across larger samples, we expanded our analysis to residences fitted with five additional sensors during the cooling season (Section 3.3.3). We identified the FFPs, which are the periods between 12 pm and 5 pm when the outdoor temperature exceeds the indoor temperature for at least an hour. These intervals reflect typical DR events as the HVAC system remains off for substantial durations. However, it is worth noting that DR events happening during heatwaves could create larger discrepancies than we identified here. In each dwelling, the CDRD and temperature deviation from the initial temperature are computed. Table 1 shows the summary statistics considering all houses with a valid FFP. Fast-reacting rooms (FRRs) and slow-reacting rooms (SRRs) are defined as rooms that achieve their CDRDs the fastest or slowest compared to other rooms in the same household, respectively. We define the *comfort gap* as a pair of numbers showing the differences between the maximum and minimum (a) CDRDs and (b) temperature deviations within a house. These results indicate that, on average, the *comfort gap* among rooms in the same house is 52 minutes and 2.37 °F. Moreover, CDRDs can be 70% longer or 40% shorter than the duration of the room with the thermostat, on average. Further, distinct rooms can deviate from the thermostat reading by an average of 48% more or 34% less. The considerably large standard deviations underscore the substantial variability and distinctive thermal behavior of each house.

The analysis in this section highlights considerable variations in the performance of thermostats during operational and DR periods with significant discrepancies from setpoint values across different rooms in a house. It becomes evident that a single thermostat does not capture the temperature complexity of an SZMRH effectively.

## 6. Evaluating the efficacy of averaging

Our second contribution is a critical evaluation of the effectiveness of averaging methodologies, conducted on our designated *Test bed 2* and *ecobee* dataset by utilizing the methodology explained in Section 3.1. While the efficacy of averaging has been investigated through simulations in Lin et al. (2002), to the best of our knowledge, no research has conducted empirical studies or utilized large datasets.

Figure 6 displays the RF of temperature deviations from the cooling setpoint for averaging and thermostat-based control in *Test bed 2*. Thermostat-based control, as depicted by the CCI values in the legend, impairs cooling comfort, with conditions worsening as one moves to higher floors (given in ascending order). Nevertheless, the zero-deviation positioning of the thermostat peak indicates that the HVAC system meets its goal. Compared to thermostat-based control, averaging resulted in higher comfort with a 45% improvement on average (excluding the basement). Yet, it is not without limitations. First, the thermostat temperature was $-6\,°F$ deviated from the setpoint, resulting in overcooling most of the time. It can also be seen that certain rooms are still less comfortable. The second room, although located in close proximity to the HVAC system, possessed numerous windows, making it prone to external effects. In upper-floor rooms, pinpointing whether the deviations on these floors arise from duct air leakage, insulation inefficiencies, or solar gain necessitates a thorough analysis. For instance, this phenomenon has been observed before in a single household and is attributed to insufficient fan power and/or duct leaks by gathering measurements with air velocity sensors (Klingensmith et al., 2014). We will try to identify such deficiencies using the *ecobee* dataset in Section 7.

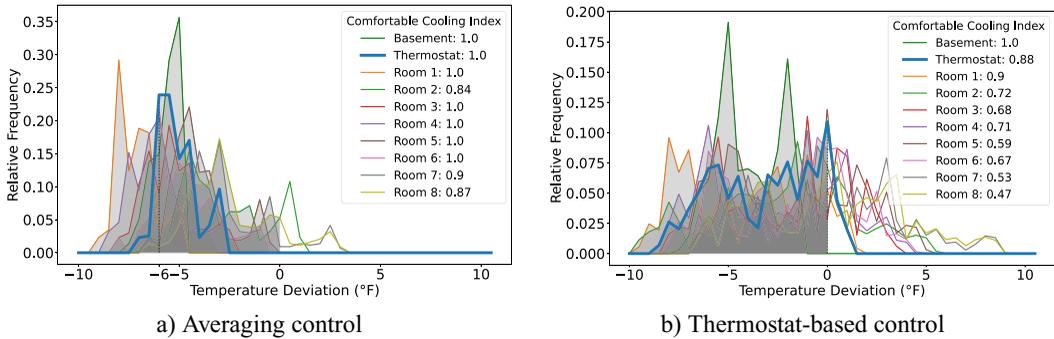

a) Averaging control                    b) Thermostat-based control

***Figure 6.*** *Relative frequencies of temperature deviations from the cooling setpoint in various rooms, represented by different colored lines. The gray area highlights the comfort zone, within which temperature fluctuations are considered comfortable for occupants. The CCI values, detailed in the legend for each respective room, quantify the proportion of time temperatures were maintained within the comfort zone.*

The *ecobee* dataset explained in Section 3.3.3 is also leveraged to understand the efficacy of averaging. The configuration of sensors, as shown in Figure 7, provides insight into the operation of houses with additional sensors (counts of which are given in the legend). During the cooling season, thermostat temperatures tend to register on the lower end of the spectrum in comparison with other sensors. This observation could be attributed to common practices of thermostat installation, such as closer proximity to air vents or the air conditioning unit than the upper floors, contributing to lower readings in the computation of average indoor temperature. Conversely, additional sensors generally record higher temperatures, possibly due to being situated further from the air conditioning unit, thereby receiving less cooled air, or due to exposure to higher levels of solar radiation during the day. Analyzing the heating season, a lesser degree of deviation is noticeable compared to the cooling season. For all households, the median temperature aligns with zero deviation from the indoor heating setpoint, indicating that all rooms achieve this reference temperature half the time. Yet, additional sensors typically report similar or lower temperatures than the thermostat reading, aligning with our previous assumption of thermostats being situated closer to the HVAC system. From this analysis, we infer that temperature variations pose a less significant issue during the heating season compared to the cooling season. Potential explanations for this phenomenon include the superior performance of HVAC systems in heating mode, the stack effect causing hot air to rise to the upper floors, and the contribution of internal and solar heat gains in maintaining room temperatures.

A limitation of the preceding methodology is that the aggregated temperature deviations might mask the unique differences among individual houses. Therefore, we need to conduct an additional analysis focusing on the hottest and coldest rooms within each house. For this particular analysis, we restrict our study to houses equipped with five additional sensors to attain a greater degree of granularity in temperature variations. Our analysis is based on the CCI values of the thermostat, the room with the lowest performance, and the room with the highest performance within each house. As illustrated in Table 2, on average, in each house, one room performs 15% worse than the room where the thermostat is located, while another room is 7% more comfortable.

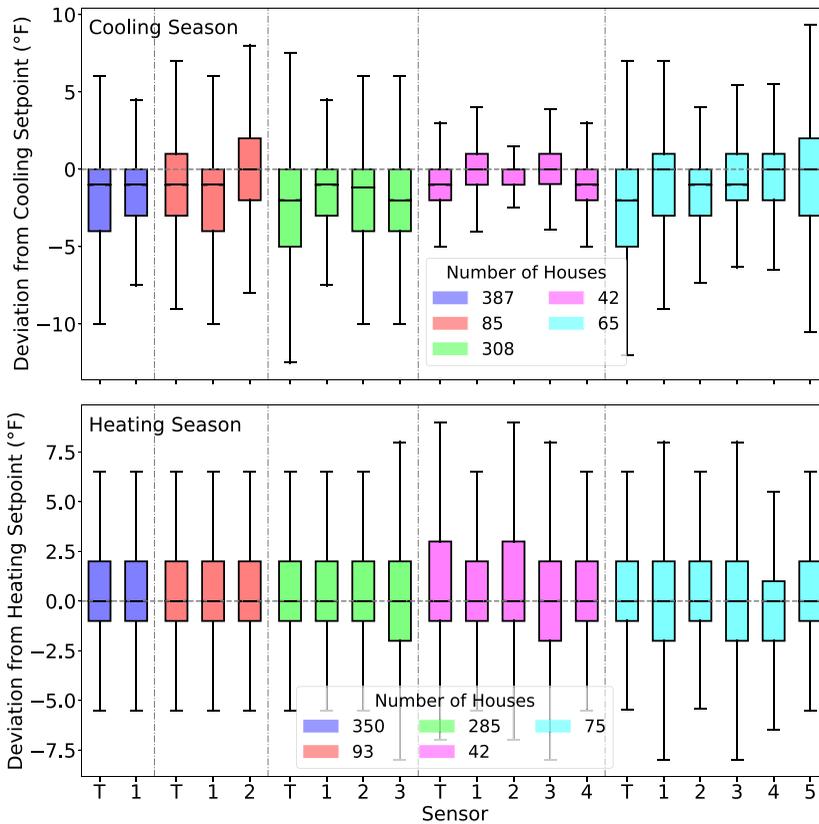

***Figure 7.*** *Distribution of temperature deviations from setpoints in houses with varied sensor counts. The top graph represents the cooling season and the bottom graph the heating season, with deviations measured in degrees Fahrenheit from respective setpoints. On the x-axis, "T" indicates measurements from the thermostat, while "1," "2," "3," "4," and "5" denote readings from the first to the fifth remote sensor, respectively. The boxplots are color-coded to reflect different groups of houses, segmented by the total number of sensors they contain. The legend shows the number of houses in each sensor group.*

Table 3 presents summary statistics of temperature deviations from both the setpoint and the average temperature (i.e., control temperature) for the coldest and hottest rooms. Deviation from the setpoint reveals the challenge the HVAC system encounters in keeping the rooms closer to the setpoint. However, it does not explain what the target of the control mechanism was for the extreme rooms. On the other hand, deviation from the control temperature signifies that even at the targeted temperature, rooms are anticipated to exhibit a certain degree of deviation. Consequently, even with a perfectly functioning HVAC system, it is expected that the coldest and hottest rooms will exhibit a deviation range of approximately $-3$ °F to 2.5 °F from the control temperature across both seasons.

## 7. Diagnosing the limitations

As the last contribution, we diagnose the reasons behind these deviations using a parameter identification methodology on up to 100 houses, each equipped with five additional sensors.

***Table 2.*** *CCI values for houses with 5 additional sensors for different room types*

|  | Mean | Median | Std. dev. | Min | Max |
|---|---|---|---|---|---|
| Thermostat room comfort | 0.92 | 1.00 | 0.17 | 0.10 | 1.00 |
| Lowest room comfort | 0.78 | 0.99 | 0.29 | 0.00 | 1.00 |
| Highest room comfort | 0.98 | 1.00 | 0.07 | 0.56 | 1.00 |

***Table 3.*** *Summary of room deviations from setpoint and average temperature*

|  | Cooling season | | Heating season | |
|---|---|---|---|---|
|  | Average | Std. dev. | Average | Std. dev. |
| **Deviations from setpoint (°F)** | | | | |
| Coldest room | −3.84 | 4.88 | −3.09 | 5.12 |
| Hottest room | 1.20 | 5.16 | 3.00 | 3.11 |
| **Deviations from average (°F)** | | | | |
| Coldest room | −2.57 | 2.91 | −3.57 | 4.62 |
| Hottest room | 2.48 | 3.58 | 2.51 | 1.97 |

### *7.1. Building thermal parameter identification*

#### *7.1.1. TTC (RC) identification*

Utilizing the FFPs according to Section 3.2.1, we inferred the $RC$ values for both cooling and heating seasons for each sensor-house pair. The subsequent process of outlier removal allowed us to infer suitable FFPs for 78.9% and 87.6% of the house-sensor pairs in the heating and cooling seasons, respectively. The increased success rate of thermal parameter identification during the cooling season could be attributed to the added parameter $RQ$, which introduces further flexibility to the process. The $RC$ values we obtained show a good alignment with those reported in previous research (John et al., 2018; Baasch et al., 2019; Wang et al., 2021). This high rate of identification in our study is likely a result of our methodological choice to extract cooling and heating data over a nine-month period (Section 3.3.3). To display the benefit of considering a longer period, we replicated the cooling season analysis using a three-month summer period and found that utilizing a longer period improved the total duration of FFPs by 84%, which in turn resulted in a 21% improvement in the percentage of identified parameters.

The distribution of $RC$ values for heating and cooling seasons is displayed in Figure 8. Each subfigure consists of two plots: the upper plot shows the collective distribution of $RC$ values, while the lower plot presents the variances in room behavior within individual houses, color-coded by state. For instance, Figure 8a reveals an unimodal distribution of $RC$ values, ranging from 1 to 27 hours for the cooling season. Contrarily, Figure 8b showcases a bimodal distribution for the heating season with a larger discrepancy in $RC$ values ranging from 1 to 157 hours. The corresponding boxplot highlights intra-house differences as large as 20 and 143 hours among rooms for cooling and heating seasons, respectively. Given that $RC$ values are solely dependent on the physical structure of the building, such variations between the $RC$ values for summer and winter are unexpected. Yet, a similar discrepancy was observed in a prior study, which was attributed to the disparity in behavioral changes of occupants, such as a tendency to leave windows open more frequently during summer (John et al., 2018). Additionally, the $RC$ value for the room where the thermostat is located is marked with light blue for each house. If the room where the thermostat is located were to accurately represent the thermal behavior of the rest of the house, the marker would be positioned in the middle of each boxplot with a notably small range. However, the plot clearly shows that this is often not the case.

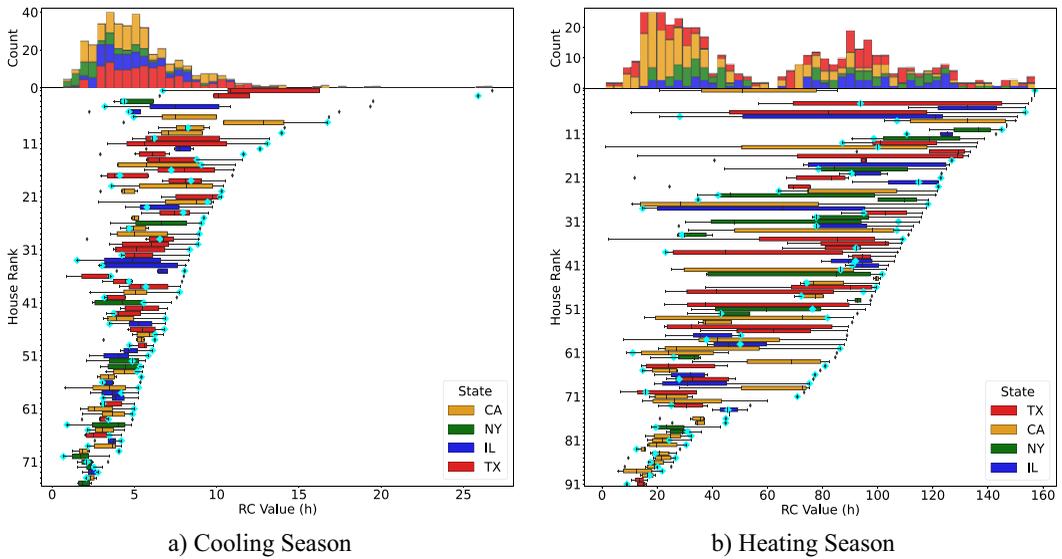

a) Cooling Season                    b) Heating Season

***Figure 8.*** *This histogram depicts the collective distribution of RC values (top), accompanied by boxplots for individual room distributions (bottom) for cooling and heating seasons. Markers indicated in light blue represent the RC values for the room where the thermostat is located.*

In the heating season, there is a 30% chance that another room in the same house will have an *RC* value twice as high as the room with the lowest *RC* value. This probability slightly decreases to 24% during the cooling season. Furthermore, our analysis revealed that more than 50% of the time, a difference of 4 hours and 54 hours is observed in the *RC* values among rooms for the cooling and heating seasons, respectively.

### 7.1.2. RQ Identification

The identification of thermal *RC* values during the cooling season concurrently produces *RQ* values. The distribution of *RQ* values can be seen in Figure 9a. Upon conducting outlier and error filtering, 87.2% of the *RQ* values were successfully inferred. Our study reveals differences as significant as 3.2 °F among rooms within the same house (see Figure 9a). According to our findings, there is a 38% probability that one room will endure twice the solar heat gain compared to another room in the same house. Furthermore, there is a 25% likelihood that one room will experience three times the solar heat gain of its counterparts within the same dwelling. Despite lacking a concrete physical interpretation, these *RQ* values provide insightful data for retrofitting decisions, especially for identifying rooms that are most susceptible to high solar irradiance. Ultimately, the varied positions of the markers suggest that rooms with thermostats may experience notably less or more solar gain compared to other rooms within the same household. This wide range of outcomes underscores that thermostat measurements may inaccurately estimate the degree of solar gain across the household, with no discernible bias toward either lower or upper limits.

### 7.1.3. RK identification

The process of estimating *RK* values for each house-sensor pair, as detailed in Section 3.2.2, commenced with the identification of nights conforming to specific constraints. The balance point model, encapsulated in Equation (3.4), was utilized to infer the *RK* values. This resulted in the *RK* value distribution ranging from 26 to 204 °F, presented in the top plot of Figure 9b with a successful identification rate of 55.9%. The boxplot in Figure 9b illuminates the considerable variability in the *RK* values across different sensors within the same house, reflecting the operational characteristics of the HVAC systems. The variation spans a broad spectrum: while some houses exhibit marginal differences, others reach deviations of approximately 80 °F. The analysis suggests that in a quarter of the instances, a room within a house receives 20% more heating input compared to the room with the least heating input. This disparity underscores the room-level variations in heating distribution, potentially signifying areas of improvement in HVAC system operation. Subsequent decisions regarding retrofitting could be informed by these insights, such as the need to enhance duct sealing to mitigate air leakage or to clean ducts to facilitate better airflow.

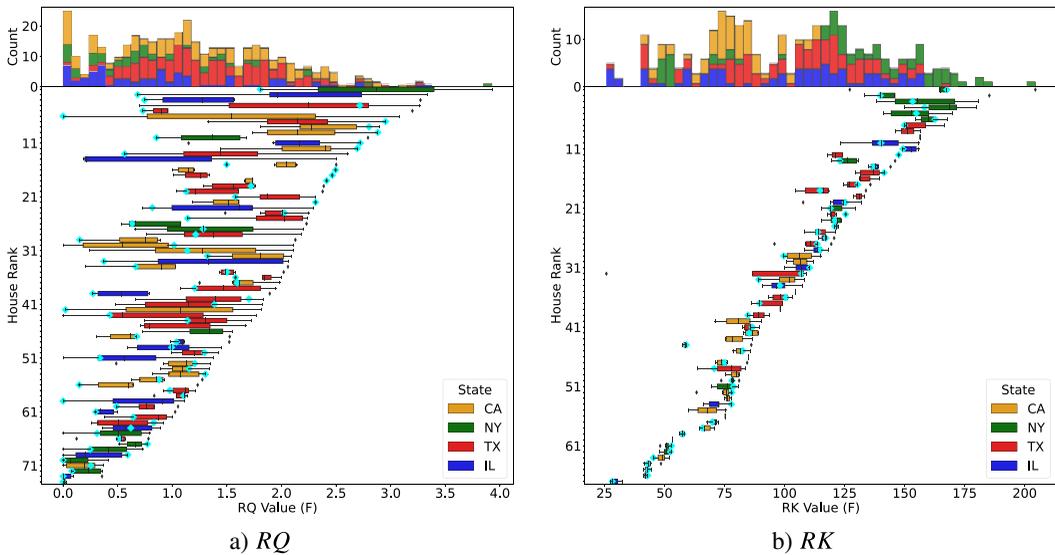

**Figure 9.** *This histogram depicts the collective distribution of RQ and RK values (top), accompanied by boxplots for individual room distributions (bottom). Markers indicated in light blue represent the RK values for the room where the thermostat is located.*

### 7.2. Interpretation of the parameters

After obtaining the $RC$, $RK$, and $RQ$ values for each house, we carried out an independent analysis to identify homes with rooms that are at a high risk of having Poor Insulation, Low Solar Gain, High Solar Gain, and Low Heating Input. All comparisons were made based on the performance of other rooms within the same house. In order to identify rooms with deficiencies, we set certain assumptions. Rooms with $RQ$ values one standard deviation below or above their mean are expected to face Low Solar Gain and High Solar Gain, respectively. Rooms exhibiting an $RK$ value one standard deviation below the mean of the house are assumed to have Low Heating Input. Poor Insulation is first detected by analyzing $RC$ data for both heating and cooling seasons by incorporating values that sit below one standard deviation from their mean. We then observed fewer houses with multiple rooms encountering Poor Insulation in the cooling season, whereas Poor Insulation from the heating season is more prevalent in houses but affects fewer rooms. This observation aligns with our initial assumption of lower $RC$ values in the cooling season due to behaviors such as occupants leaving windows open more frequently. As a result, to accurately represent the actual physical attributes of the house, and not merely occupant behavior-related deficiencies, we define Poor Insulation using $RC$ values from the heating season. It should be noted that since there are no studies that utilized the ST data for such purposes, there is no accepted metric for us to use. However, our approach is akin to the smart meter data-based inefficient house detection methodology conducted by Iyengar et al. (2016). While this limited approach does not conclusively indicate such inefficiencies, it does help identify rooms with a high risk of such deficiencies.

Our observations indicate that a maximum of two rooms within a house exhibit such deficiencies. Although the actual number may be higher, this is a plausible estimate considering the constraints of having only six sensor readings from each house. Table 4 presents the number of houses exhibiting these deficiencies in just one room, two rooms, and the total. The percentages have been calculated based on the total data available for each instance separately, which explains why the denominator varies for each row. Overall, 96% of the houses exhibit High Solar Gain, with 78% in one room and 18% in two rooms. Low Heating Input emerges as the second most common deficiency, occurring in 85% of the houses overall, with 70% of the instances being confined to just one room. Low Heating Input is recorded in 78% of the houses, 69% of which observe it in only one room. Lastly, 70% of the houses are at risk of Poor Insulation, with 15% experiencing it in two rooms.

***Table 4.** Number of houses with deficiencies*

| Number of houses | 1 room | 2 rooms | Total (1 or 2 rooms) |
| --- | --- | --- | --- |
| Low solar gain | 50 (69.44%) | 6 (8.33%) | 56 (77.77%) |
| High solar gain | 56 (77.77%) | 13 (18.05%) | 69 (95.83%) |
| Low heating input | 38 (70.37%) | 8 (14.81%) | 46 (85.18%) |
| Poor insulation | 45 (54.87%) | 12 (14.63%) | 57 (69.51%) |

***Table 5.** Statistical analysis of the thermal parameter difference of rooms from the thermostat*

|  | Mean | Median | Std. dev. | Min | Max | Coeff. of var. (%) |
| --- | --- | --- | --- | --- | --- | --- |
| *RC* (h) | −8.24 | −1.09 | 32.79 | −108.95 | 79.29 | 398 |
| *RK* (°F) | −0.30 | −0.21 | 7.10 | −26.74 | 14.78 | 2367 |
| *RQ* (°F) | 0.08 | 0.12 | 0.53 | −1.39 | 1.20 | 662 |

In our final analysis, we examined the assumption that thermostat placements—ideally unaffected by vents or solar radiation—accurately represent overall house conditions. We computed the differences in thermal parameters across rooms from the room where the thermostat is located, with the summary statistics presented in Table 5. While the relatively low mean/median values might seem to initially endorse this assumption, the high standard deviations (shown by the coefficient of variance) indicate significant variation in each home's thermostat-related behavior. This prevents us from drawing robust conclusions regarding the thermal conditions experienced by thermostats due to their physical locations. However, we can assert that they do not adequately represent the rest of the house.

## 8. Energy implications of remote sensors

While the addition of more sensors in homes, as discussed in Section 6, has the potential to enhance comfort levels, their impact on energy consumption remains uncertain. These sensors could lead to energy savings by enabling heating and cooling systems to operate based on occupancy (either through motion sensors or schedules), thereby optimizing energy use. However, they could also increase energy consumption since shifts in room occupancy can lead to sudden changes in the control temperatures set by the system. To investigate this, we begin with an exploratory data analysis aimed at identifying correlations between various attributes and the energy consumption of households. This step is crucial for isolating the effects of other variables. Subsequently, we delve into a comprehensive statistical analysis to gauge the influence of sensor quantity on energy consumption.

### 8.1. Explanatory data analysis

In our exploratory analysis, we augmented the *ecobee* DYD dataset with additional information from *ecobee* metadata, incorporating attributes like floor area and number of occupants. Although the *ecobee*

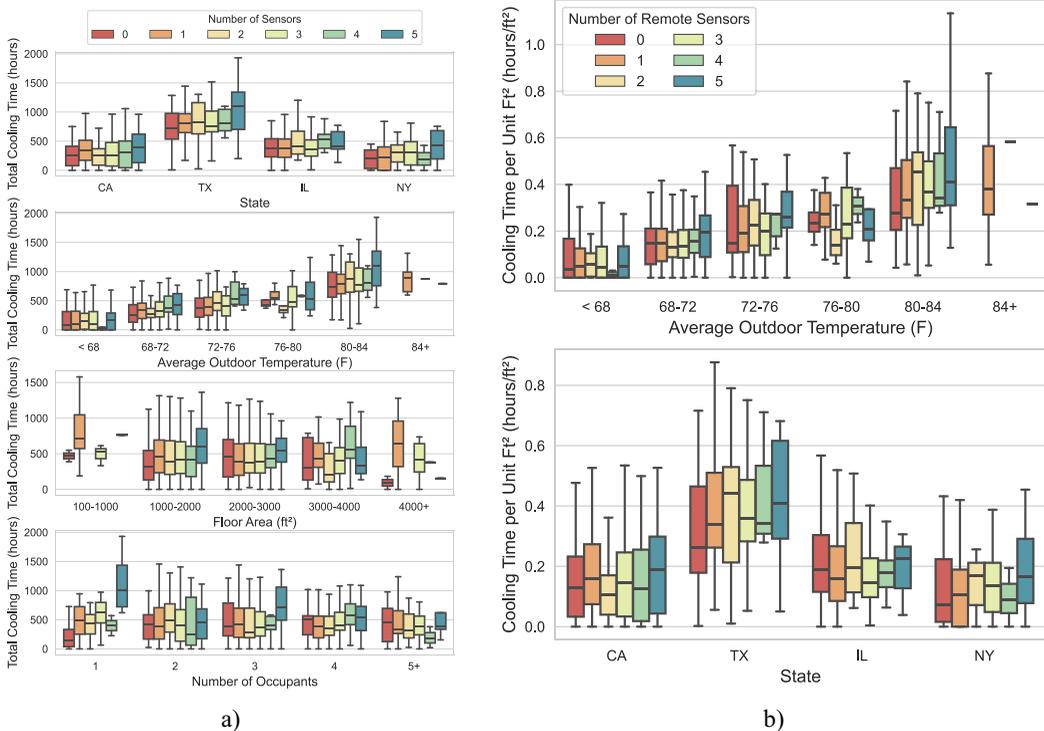

a)                                          b)

***Figure 10.*** *Panel (a) depicts box plots that categorize total cooling time by the number of remote sensors (0–5) across various household attributes such as state locations, average outdoor temperatures, floor area, and occupancy. Panel (b) presents cooling time per unit floor area as box plots, stratified by the same states and average outdoor temperatures, revealing how cooling efficiency correlates with the presence of remote sensors within the dwelling.*

DYD dataset lacks direct measurements of energy consumption, it includes the run time of cooling equipment as a feature. By aggregating the cooling run times for homes equipped with multi-stage cooling systems, we derived a consolidated feature termed "Cooling Run Time." This metric serves as an indirect indicator of energy consumption during the cooling season. However, it is worth noting that this metric, while useful for estimating energy consumption, does not account for the varying energy usage rates associated with different operational stages in multi-stage systems. Figure 10a shows the distribution of the total cooling run time of houses for the summer season, colored by the number of sensors they have. As expected, we see a clear proportional relationship between the number of sensors and the total cooling time when the effect of the average outdoor temperature is removed. In terms of states, we see that only Texas has a similar trend to the one we see in outdoor temperature. However, when data are segmented by either floor area or number of occupants, the increase in sensor count does not significantly impact energy consumption, suggesting that these factors might play a minimal role in the observed energy usage patterns.

Acknowledging the potential for the binning of floor area to obscure its impact, we embarked on a secondary analysis. This involved the introduction of a new attribute, cooling run time per square foot, derived by dividing a house's total cooling time by its floor area. Figure 10b presents this metric's effects across different categorizations, namely, state and average outdoor temperature. This deeper dive reveals that, absent state-specific effects, there is no evident trend toward increased energy use. However, the analysis suggests an increase in energy consumption associated with a greater number of sensors when considering the intersection with average outdoor temperatures.

### 8.2. Statistical analysis

The exploratory data analysis has provided initial insights into the potential impact of additional sensors on energy consumption. Building upon these findings, we now aim to delve deeper by conducting a comprehensive statistical analysis. This offers empirical evidence concerning the correlation between sensor deployment and energy usage. Our methodology is inspired by the approach used by Barreca et al. (2016), which explored the effects of temperature on mortality rates, adapting it to our context of energy consumption in residential settings.

To commence, the year-long dataset from *ecobee*'s DYD program was processed to obtain hourly data points. This resampling involved summing up the run time of both cooling and heating equipment for each hour, while temperature readings were averaged over the same period. Further, we computed a *combined* run time by summing cooling and heating run times. This allowed us to make claims related to the general energy consumption of the houses. We then calculated the Duty Cycle (unitless) for both cooling, heating, and combined run times by dividing the equipment's run time (in seconds) by 3600 seconds, thus obtaining a ratio reflecting the proportion of time the HVAC system was operational within each hour. Additionally, the number of sensors at any given hour was determined by counting the number of non-null temperature readings from the remote sensors.

We have formulated our regression problem as follows:

$$DC_{it} = \sum_{j=0}^{5} \left( \beta_j \cdot s_{j,it} \times T_{it} \right) + \alpha_i + \rho_h + v_m + \epsilon_{it}, \tag{8.1}$$

where $DC_{it}$ represents the heating, cooling, or *combined* duty cycle for house $i$ at hour$t$.

$s_{j,it}$ represents the presence of sensor count $j$ for house $i$ at hour $t$.

$T_{ti}$ represents the outdoor temperature for house $i$ at hour $t$.

$\beta_j$ are coefficients for the interaction of $j^{th}$ additional sensor dummy (as $j$ goes from 0 to 5) with outdoor temperature $T_{it}$ at hour $t$.

$\alpha_i$ represents the fixed effects for each house, controlling for unobserved heterogeneity across houses.

$\rho_h$ represents the fixed effects for each hour $h$ of the day, controlling for diurnal variations.

$v_m$ represents the fixed effects for each month $m$ of the sample, controlling for time-varying patterns across the dataset.

$\epsilon_{it}$ is the error term for house $i$ at hour $t$.

The R environment is used to fit the models. The summary of the results of the model fit for the *combined* duty cycle can be found in Table 6. Our initial analysis on *combined* duty cycle demonstrates statistically significant results for each sensor's impact on energy consumption. All coefficients being positive also implies that cooling is the dominant energy consumption. Furthermore, since remote sensors are often sold in pairs, our analysis indicates that users enhancing their *ecobee* premium system with a two-sensor pack can expect a 4.9% rise in energy usage per degree Fahrenheit change in outdoor temperature. Progressing from no extra sensors to a total of 5 can escalate cooling consumption by 17.8%.

Figure 11 depicts the impact of a 1 °F increase in outdoor temperature on the total energy consumption, with brackets showing the 95% confidence intervals. The trend indicates that increasing the number of sensors results in increased energy consumption per degree change in the outdoor temperature. The largest jump with the addition of a sensor happens when switching from 0 to 1 with a 6.5% increase in demand. This finding gains significance with the introduction of newer *ecobee* thermostat models like the *ecobee* premium, which comes with an additional sensor. Furthermore, since remote sensors are sold in pairs, our analysis indicates that users enhancing their *ecobee* premium system with a two-sensor pack can expect a 4.9% rise in their energy usage per degree Fahrenheit change in outdoor temperature. Moving from no extra sensors to a total of 5 can escalate consumption by 17.8%.

The summary of the results of the two model fit for the heating and cooling duty cycle can be found in Table 7. For cooling, data clearly show a positive relationship between the number of sensors and energy consumption, with all coefficients from $\beta_0$ through $\beta_5$ being positive and statistically significant. This is an expected result since the increase in outdoor temperature would result in an increase in cooling energy

**Table 6.** *Statistical summary of the model evaluating the effect of sensor count and outdoor temperature on HVAC (combined) duty cycle (1/°F)*

| Term | Estimate | Std. error | *t* value | Pr(>|t|)[a] |
|------|----------|-----------|-----------|------------|
| $\beta_0$ | 0.0128 | 0.000262 | 48.9 | < 2.2e–16 *** |
| $\beta_1$ | 0.0136 | 0.000252 | 54.2 | < 2.2e–16 *** |
| $\beta_2$ | 0.0141 | 0.000261 | 54.2 | < 2.2e–16 *** |
| $\beta_3$ | 0.0143 | 0.000265 | 54.0 | < 2.2e–16 *** |
| $\beta_4$ | 0.0146 | 0.000302 | 48.4 | < 2.2e–16 *** |
| $\beta_5$ | 0.0151 | 0.000356 | 42.4 | < 2.2e–16 *** |

*Note.* Fixed-effects: id: 1000, hour: 24, and month: 12. Standard errors: clustered (id).
[a]Signif. codes: 0'***' 0.001'**' 0.01'*' 0.05'.' 0.1' 1.

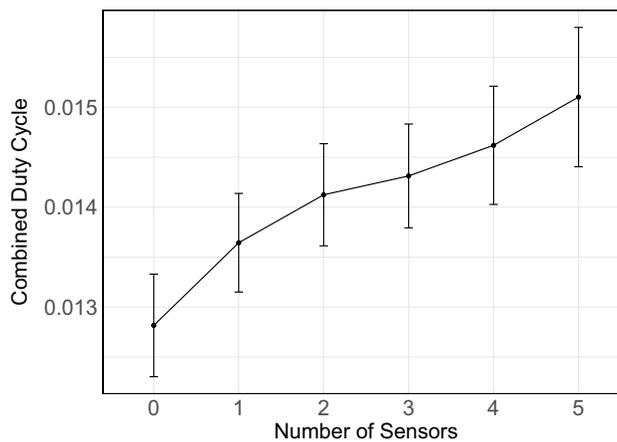

**Figure 11.** *Estimated impact of 1 °F in outdoor temperature on HVAC (combined) duty cycle. 95% confidence interval is shown using brackets.*

consumption. Notably, the magnitude of these coefficients increases with each additional sensor, indicating that the addition of each sensor contributes to higher energy use. This consistent increase across sensor counts strongly suggests that the deployment of more sensors leads to significant increases in cooling energy consumption, likely due to the sensors capturing more spatial temperature variability and triggering more frequent or intense cooling responses.

In contrast, the heating duty cycle presents a more complex picture. The coefficients for the number of sensors interacting with outdoor temperature show negative estimates with statistical significance. Negative values are expected due to the inverse relation between outdoor temperature and heating run time. This mixed outcome suggests a nuanced relationship where, under certain outdoor temperature conditions, increasing the number of sensors does not uniformly increase energy consumption for heating. In some cases, it might even reduce it, possibly due to optimized heating strategies facilitated by the data from these sensors or the stack effect helping with the rooms on upper floors.

The comparison of values in Table 7 shows that the coefficients for heating are significantly smaller, likely because there's less need for heating over a narrower temperature range characteristic of warmer climates. Therefore, the model tends to identify smaller coefficients for heating consumption. This effect on the coefficients wouldn't be present if the relationship between outdoor temperature and energy consumption were linear. However, given the exponential nature of this relationship, the temperature differential within warmer ranges of outdoor temperature leads to substantially smaller heating duty cycles. This outcome is anticipated, considering the dataset includes a significant number of houses from California, where cooling is often the primary energy demand due to the state's warmer climate.

**Table 7.** *Statistical summary of the model evaluating the effect of sensor count and outdoor temperature on cooling and heating duty cycle (1/°F)*

| | Cooling | | | | Heating | | | |
|---|---|---|---|---|---|---|---|---|
| Term | Estimate | Std. error | *t* value | Pr(>\|t\|)[a] | Estimate | Std. error | *t* value | Pr(>\|t\|)[a] |
| $\beta_0$ | 0.0129 | 0.000262 | 49.4 | < 2.2e–16 *** | −0.0000960 | 0.0000112 | −8.58 | < 2.2e–16 *** |
| $\beta_1$ | 0.0137 | 0.000252 | 54.6 | < 2.2e–16 *** | −0.0000920 | 0.00000994 | −9.27 | < 2.2e–16 *** |
| $\beta_2$ | 0.0142 | 0.000260 | 54.8 | < 2.2e–16 *** | −0.000105 | 0.0000128 | −8.27 | 4.31e–16 *** |
| $\beta_3$ | 0.0144 | 0.000264 | 54.5 | < 2.2e–16 *** | −0.0000790 | 0.00000974 | −8.10 | 1.56e–15 *** |
| $\beta_4$ | 0.0147 | 0.000303 | 48.5 | < 2.2e–16 *** | −0.0000940 | 0.0000154 | −6.11 | 1.47e–09 *** |
| $\beta_5$ | 0.0152 | 0.000363 | 41.9 | < 2.2e–16 *** | −0.000114 | 0.0000265 | −4.32 | 1.74e–05 *** |

*Note.* Fixed-effects: id: 1000, hour: 24, and month: 12. Standard errors: clustered (id).
[a]Signif. codes: 0 '***' 0.001 '**' 0.01 '*' 0.05 '.' 0.1 '' 1.

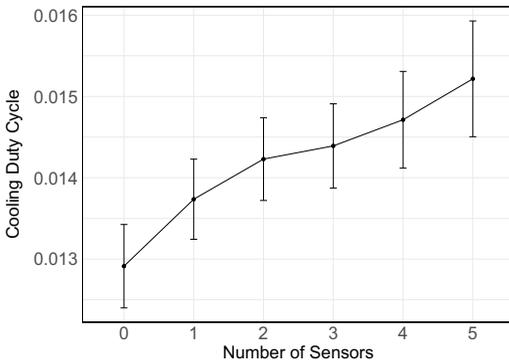

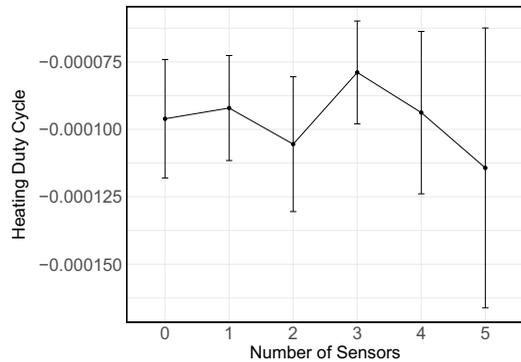

a) Impact on Cooling Duty Cycle                b) Impact on Heating Duty Cycle

**Figure 12.** *Estimated impact of 1 °F in outdoor temperature on (a) cooling and (b) heating duty cycle. 95% confidence interval is shown using brackets.*

Figure 12 illustrates the influence of a 1 °F increase in outdoor temperature on the cooling (a) and heating (b) duty cycles, with brackets indicating the 95% confidence intervals. For cooling, the trend is almost identical to *combined* duty cycle, reinforcing the notion that cooling predominantly drives overall energy consumption. There is a discernible positive trend as the number of sensors increases. Notably, the transition from 0 to 1 sensor incurs a 6.4% uptick in energy consumption per degree change in outdoor temperature. Furthermore, we observe that users augmenting their *ecobee* premium with a two-sensor pack can cause a 4.8% rise in cooling energy usage per degree change in outdoor temperature. Lastly, the switch from no extra sensors to a total of 5 can surge the cooling consumption by 17.8%, which is identical to what was observed for *combined* energy consumption.

When switched to heating, a fluctuating correlation happens, as can be seen from 12(b). Generally, the coefficients decrease as the number of sensors grows. It is important to note that smaller negative coefficients indicate a larger reduction from the baseline consumption, leading to smaller energy consumption. We observe that having one additional sensor could reduce the energy consumption by 4.2% per outdoor degree change. Additionally, augmenting the existing *ecobee* premium thermostat (that already comes with an extra sensor) with two additional sensors can lead to a 14.1% decrease in heating demand. Surprisingly, we see that adding an additional sensor could result in a 25% reduction (from 2 to 3) or a 19% increase (from 3 to 4) in heating energy consumption. Lastly, the switch from not having any sensors to having 5 results in an 18.8% increase.

**Table 8.** *Statistical summary of the model evaluating the effect of sensor count, outdoor temperature, and seasonality on cooling and heating duty cycle (1/°F)*

| Term | Cooling | | | | Heating | | | |
|---|---|---|---|---|---|---|---|---|
| | Estimate | Std. error | $t$ value | Pr(>\|t\|)[a] | Estimate | Std. error | $t$ value | Pr(>\|t\|)[a] |
| $\beta_{0:Spring}$ | 0.00910 | 0.000308 | 29.6 | < 2.2e–16 *** | −0.000215 | 0.0000200 | −11.0 | < 2.2e–16 *** |
| $\beta_{0:Summer}$ | 0.0158 | 0.000331 | 47.7 | < 2.2e–16 *** | 0.0000390 | 0.0000110 | 3.47 | 5.44e–04 *** |
| $\beta_{0:Autumn}$ | 0.0118 | 0.000271 | 43.6 | < 2.2e–16 *** | −0.0000950 | 0.0000120 | −7.91 | 6.83e–15 *** |
| $\beta_{0:Winter}$ | 0.00249 | 0.000416 | 6.00 | 2.92e–09 *** | −0.00127 | 0.0000750 | −16.9 | < 2.2e–16 *** |
| $\beta_{1:Spring}$ | 0.00985 | 0.000302 | 32.6 | < 2.2e–16 *** | −0.000203 | 0.0000170 | −11.6 | < 2.2e–16 *** |
| $\beta_{1:Summer}$ | 0.0167 | 0.000308 | 54.3 | < 2.2e–16 *** | 0.0000480 | 0.0000100 | 4.80 | 1.86e–06 *** |
| $\beta_{1:Autumn}$ | 0.0126 | 0.000258 | 48.9 | < 2.2e–16 *** | −0.0000920 | 0.0000150 | −6.27 | 5.49e–10 *** |
| $\beta_{1:Winter}$ | 0.00294 | 0.000404 | 7.26 | 7.58e–13 *** | −0.00133 | 0.0000720 | −18.4 | < 2.2e–16 *** |
| $\beta_{2:Spring}$ | 0.0105 | 0.000307 | 34.2 | < 2.2e–16 *** | −0.000202 | 0.0000200 | −10.3 | < 2.2e–16 *** |
| $\beta_{2:Summer}$ | 0.0171 | 0.000332 | 51.5 | < 2.2e–16 *** | 0.0000330 | 0.0000120 | 2.68 | 7.47e–03 ** |
| $\beta_{2:Autumn}$ | 0.0131 | 0.000269 | 48.8 | < 2.2e–16 *** | −0.000105 | 0.0000160 | −6.74 | 2.63e–11 *** |
| $\beta_{2:Winter}$ | 0.00369 | 0.000404 | 9.15 | < 2.2e–16 *** | −0.00133 | 0.0000750 | −17.8 | < 2.2e–16 *** |
| $\beta_{3:Spring}$ | 0.0105 | 0.000306 | 34.3 | < 2.2e–16 *** | −0.000193 | 0.0000170 | −11.7 | < 2.2e–16 *** |
| $\beta_{3:Summer}$ | 0.0173 | 0.000324 | 53.5 | < 2.2e–16 *** | 0.0000600 | 0.0000120 | 5.04 | 5.57e–07 *** |
| $\beta_{3:Autumn}$ | 0.0132 | 0.000269 | 49.2 | < 2.2e–16 *** | −0.0000830 | 0.0000140 | −5.90 | 5.02e–09 *** |
| $\beta_{3:Winter}$ | 0.00363 | 0.000393 | 9.23 | < 2.2e–16 *** | −0.00128 | 0.0000690 | −18.6 | < 2.2e–16 *** |
| $\beta_{4:Spring}$ | 0.0108 | 0.000351 | 30.7 | < 2.2e–16 *** | −0.000216 | 0.0000220 | −9.66 | < 2.2e–16 *** |
| $\beta_{4:Summer}$ | 0.0178 | 0.000356 | 49.9 | < 2.2e–16 *** | 0.0000490 | 0.0000150 | 3.27 | 1.10e–03 ** |
| $\beta_{4:Autumn}$ | 0.0134 | 0.000340 | 39.4 | < 2.2e–16 *** | −0.000102 | 0.0000190 | −5.46 | 6.03e–08 *** |
| $\beta_{4:Winter}$ | 0.00400 | 0.000431 | 9.28 | < 2.2e–16 *** | −0.00126 | 0.0000810 | −15.7 | < 2.2e–16 *** |
| $\beta_{5:Spring}$ | 0.0112 | 0.000396 | 28.3 | < 2.2e–16 *** | −0.000231 | 0.0000330 | −6.93 | 7.50e–12 *** |
| $\beta_{5:Summer}$ | 0.0183 | 0.000434 | 42.1 | < 2.2e–16 *** | 0.0000190 | 0.0000270 | 0.701 | 0.483 |
| $\beta_{5:Autumn}$ | 0.0137 | 0.000329 | 41.7 | < 2.2e–16 *** | −0.000114 | 0.0000260 | −4.39 | 1.24e–05 *** |
| $\beta_{5:Winter}$ | 0.00392 | 0.000430 | 9.10 | < 2.2e–16 *** | −0.00131 | 0.0000810 | −16.1 | < 2.2e–16 *** |

*Note.* Fixed-effects: id: 1000, hour: 24, and month: 12. Standard errors: clustered (id).
[a]Signif. codes: 0 '***' 0.001 '**' 0.01 '*' 0.05 '.' 0.1 ' ' 1.

Recognizing the potential to uncover insights into how the impact of the number of sensors on energy consumption changes over time, we have identified seasonality as a critical factor for our analysis. This choice is driven by the hypothesis that the efficacy of sensors in influencing energy consumption could vary significantly across different seasons, due to variations in environmental conditions and heating or cooling needs.

To thoroughly investigate the influence of seasonality, we have reformulated our regression model to incorporate interactions between the number of sensors, outdoor air temperature, and the seasons. This revised approach allows us to examine how seasonal changes affect the relationship between sensor deployment and energy consumption.

$$DC_{it} = \sum_{k=1}^{4} \sum_{j=0}^{5} \left( \beta_{jk} \cdot SD_k \times s_{j,it} \times T_{it} \right) + \alpha_i + \rho_h + \nu_m + \epsilon_{it}. \tag{8.2}$$

In the updated equation, denoted as Equation (8.2), two primary modifications have been made compared to Equation (8.1). First, we introduced $SD_k$, representing the dummy variable for the $k^{th}$ season, to our model. Second, the dummy variables for the number of sensors now encompass the scenario of having zero additional sensors, which previously served as the baseline in Equation (8.1). The inclusion of the zero additional sensor condition in our current analysis aims to elucidate the seasonal effects on scenarios where no extra sensors are used.

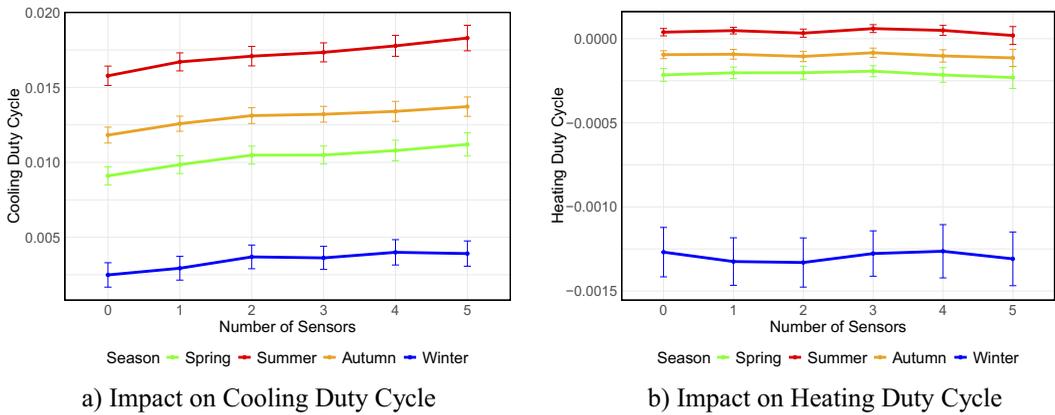

a) Impact on Cooling Duty Cycle          b) Impact on Heating Duty Cycle

***Figure 13.*** *Estimated impact of 1 °F in outdoor temperature on (a) cooling and (b) heating duty cycle for each season. 95% confidence interval is shown using brackets.*

The outcomes derived from Equation (8.2) are detailed in Table 8. Notably, with the exception of $\beta_{5:Summer}$ in the heating analysis, all results exhibit statistical significance. This underscores a clear correlation between the number of sensors and heating usage, particularly when the influence of seasonality is considered.

Figure 13 demonstrates the estimated impact of a 1 °F increase in outdoor temperature on a) cooling and b) heating duty cycles across different seasons. For cooling energy consumption, our results align with anticipated trends, revealing a pronounced increase in cooling run time during the summer months. The distinction between autumn and spring is minimal, with spring presenting slightly lower cooling usage. Winter sees the smallest effect, yet the overall ascending trend observed in Figure 12a persists, especially during summer, when cooling demand rises with each additional sensor.

The heating analysis presents a more variable pattern. In this context, a downward shift in heating duty cycle values indicates a decrease in heating expenses. The analysis reveals that incorporating just one additional sensor can reduce heating equipment use by 4.4%. However, expanding an *ecobee* premium by adding a pack of remote sensors (transitioning from 1 to 3 sensors) leads to a 3.6% increase in heating expenses. Collectively, these findings suggest that the inclusion of extra sensors has a marginal impact on heating energy consumption. This aligns with prior observations regarding the smaller temperature disparities during the heating season, as depicted in Figure 7. Such phenomena were attributed to the stack effect and the roles of internal and solar heat gains in maintaining indoor temperatures. Additionally, unaccounted space heaters might have contributed to this phenomenon.

## 9. Discussion and conclusion

This study offers a comprehensive exploration of the thermal behavior variability and energy efficiency in SZMNSs using data from two residential test beds and a publicly available panel dataset and metadata from *ecobee.*

Based on our initial analysis, we observed that 67% of the houses with *ecobee* thermostats are equipped with at least one remote sensor. Moreover, 15,404 houses out of the total dataset (13.6%) have participated in DR through the eco + program. Among these, 73% have at least one sensor, which indicates the potential for incorporating additional sensory information while assessing the thermal comfort during DR.

The subsequent analysis identifies significant temperature discrepancies across rooms, often overlooked by traditional thermostats. We displayed a consistent 3 °F discrepancy between the thermostat's reading and the setpoint, with comfort indices falling below 50% within rooms. Further, we noted a significant disparity during DR events, where the duration of comfort was typically 70% longer or 40% shorter compared to the thermostat-controlled room.

Further, we underscore that while averaging may mitigate some discrepancies, substantial deviations outside comfort bounds persist. In our test bed, averaging techniques demonstrated an average of 45% improvement in cooling operations. Despite this improvement, the thermostat reading still exhibited an average deviation of $-6$ °F. On a larger scale, we found that individual rooms display an expected deviation ranging from $-3$ °F to 2.5 °F from this average (i.e., control temperature). This considerable discrepancy highlights that, even with averaging, the targeted temperatures for rooms are typically outside of the comfort bounds ($\pm 2$ °F).

Our detailed parameter identification study further illuminates the potential root causes of these deviations. In 95% of houses, we found rooms with High Solar Gain, 85% with Low Heating Input, and 70% with Poor Insulation. Moreover, we observe that common assumptions about thermostat placement are often inaccurate, as they are typically located in rooms that do not reflect the whole house's thermal conditions.

Lastly, our research indicates that cooling energy consumption per degree of outdoor temperature change increases with the addition of more sensors. On the other hand, heating usage has been observed to fluctuate from $-20\%$ to $+25\%$ with an increase in the number of sensors. Specifically, homeowners adding a pack of remote sensors to their *ecobee* premium thermostats may see a 4.8% rise in cooling energy consumption per degree Fahrenheit change in outdoor temperature, while heating consumption could decrease by 14.1%.

In conclusion, despite moderate improvements with averaging techniques, significant temperature deviations in SZMNSs still exist, affecting thermal comfort during regular operations and DR events. The availability of remote sensors paves the way for utilizing additional sensory information in evaluating services provided by buildings such as thermal comfort and DR. Energy implications of remote sensors reveal the trade-off users must pay in order to achieve more comfort, especially in cooling. The methodology used in this study can benefit future HVAC designs by enabling ST companies to leverage these algorithms for room deficiency detection, enhancing energy efficiency of remote sensors, improving retrofit cost estimates, and guiding optimal thermostat placement for accurate representation of a home's thermodynamic characteristics.

Some limitations of this study are 1) the characteristics of the *ecobee* DYD dataset, which could have influenced our findings such as its temperature resolution or the limited number of houses equipped with sensors, and 2) the metrics we employed to identify deficiencies have not yet been validated elsewhere. Despite these potential limitations, the findings remain illustrative and point to promising new research avenues. While subsequent studies might develop refined metrics, leveraging building thermal parameters for more nuanced room-level deficiency evaluations, our analysis still offers substantial insights. Future work could include an analysis of estimating overriding behavior during DR events by utilizing room-level occupancy and temperature data. Moreover, existing model predictive control methodologies could be extended to room-level granularity using the methods we deployed. Further, the effect of temperature variance in gray box modeling approaches utilizing indoor average temperature should be investigated. Lastly, the dataset's lack of counterfactuals precludes evaluating variations in comfort conditions based on different sensor counts, as each house is controlled according to its available sensors. This restriction hinders a detailed cost–benefit analysis, but future studies could explore these dimensions in a controlled environment to effectively demonstrate the trade-offs between sensor-induced comfort and their financial implications.

**Acknowledgements.** We would like to thank *ecobee* and its customers for their data. The content of this manuscript has been presented in part at the 10th ACM International Conference on Systems for Energy-Efficient Buildings, Cities, and Transportation (Mulayim and Bergés, 2023).

**Data availability statement.** Our code and data from the two test beds can be found in the following Github repository https://github.com/INFERLab/Sensors4SingleZoneSystems.

**Author contribution.** Conceptualization, methodology, and writing—review and editing: O.B.M, M.B, and E.S. Data curation, formal analysis, investigation, validation, visualization, and writing—original draft: O.B.M. Supervision: M.B and E.S. Funding acquisition: M.B. All authors approved the final submitted draft.

**Funding statement.** The authors would like to gratefully acknowledge the support provided by the Pennsylvania Infrastructure Technology Alliance (PITA).

**Competing interest.** Mario Bergés holds concurrent appointments as a Professor of Civil and Environmental Engineering at Carnegie Mellon University and as an Amazon Scholar. This paper describes work at Carnegie Mellon University and is not associated with Amazon.

**Ethical statement.** The research meets all ethical guidelines, including adherence to the legal requirements of the study country.